\input{psfig.sty}

\documentclass[preprint2]{aastex}

\begin{document}
 
\title{The Broad Band Spectrum of MXB 1728--34 Observed by BeppoSAX}

\author{T. Di Salvo\altaffilmark{1}, 
R. Iaria\altaffilmark{1},
L. Burderi\altaffilmark{2},
N. R. Robba\altaffilmark{1}}
\altaffiltext{1}{Dipartimento di Scienze Fisiche ed Astronomiche, 
Universit\`a di Palermo, via Archirafi n.36, 90123 Palermo, Italy}
\altaffiltext{2}{Osservatorio Astronomico di Roma, Via Frascati 33, 
00040 Monteporzio Catone (Roma), Italy}
  
\authoremail{disalvo@gifco.fisica.unipa.it}

\begin{abstract}
We report on the results of a broad band (0.1--100 keV) spectral analysis
of the bursting atoll source MXB 1728--34 (4U 1728--34) observed by the 
BeppoSAX satellite. 
Three bursts were present during this observation.  The spectrum during
the bursts can be fitted by a blackbody with a temperature of $\sim 2$ keV.
The radius of the blackbody emitting region is compatible with the radius of 
the neutron star if we correct for the difference between the observed
color temperature and the effective temperature. From the bursts we also 
estimate a distance to the source of $\sim 5.1$ kpc.
MXB 1728--34 was in a rather soft state during the BeppoSAX observation.
The persistent spectrum is well fitted by a continuum consisting of a
soft blackbody emission and a comptonized spectrum.  We interpreted the soft 
component as the emission from the accretion disk.  
Taking into account a spectral hardening factor of $\sim 1.7$ (because of 
electron scattering which modifies the blackbody spectrum emitted by the 
disk), we estimated that the inner disk radius is
$R_{\rm in} \sqrt{\cos i} \sim 20$ km, where $i$ is 
the inclination angle.  The comptonized component could originate
in a spherical corona, with temperature $\sim 10$ keV and optical depth 
$\sim 5$, surrounding the neutron star.  A broad gaussian emission
line at $\sim 6.7$ keV is observed in the spectrum, probably emitted in the
ionized corona or in the inner part of the disk. Another emission line is 
present at $\sim 1.66$ keV.
No reflection component is detected with high statistical significance, 
probably because of the low temperature of the corona in this state of the 
source. If the iron emission line is due to reflection of the comptonized 
spectrum by the accretion disk, it requires a ionized disk ($\xi \sim 280$)
and a solid angle of $\sim 0.2$ (in units of $2\pi$) subtended by the 
reflector as seen from the corona.
\end{abstract}


\section{Introduction}

The recent broad band spectral analysis of neutron star Low Mass X-ray 
Binaries (hereafter NS)
has shown that their spectra are often more complex than previously thought.
In fact, similarly to accreting Black Holes (BH), these spectra
could be described by a power law with high energy cutoff and a soft
excess at 0.5--1 keV. The thermal Comptonization of these soft photons 
by hot electrons probably originates the power law. Moreover an
emission line is usually present at $\sim 6.4$ keV interpreted as  
fluorescence from iron in low ionization states. 
The probable origin of this line is from the reprocessed emission from 
the accretion disk surface illuminated by the primary comptonized spectrum. 
In this case one would also expect the presence of a 
bump between 20 and 40 keV due to Compton reflection of the primary
spectrum by the disk. Indeed this reflection bump
has been observed in the spectra of some NSs (Barret et al. 1999; 
Piraino et al. 1999; Yoshida et al. 1993), 
usually with reflection amplitudes ({\it i.e.} the solid angle $\Omega/2\pi$
subtended by the reflector as seen from the corona)  
lower than 0.3. In these cases a correlation has been observed between the
photon index of the primary spectrum and the reflection amplitude of the
reprocessed component (Zdziarski et al. 1999; Barret et al. 1999; 
Piraino et al. 1999), the same observed in Seyfert galaxies and galactic
BHs. It was originally thought that the electron temperature in the 
scattering cloud is lower for the NSs than for the BHs
(Tavani \& Barret 1997; Zdziarski et al. 1998; Churazov et al. 1997).
This is in agreement with the expectation that an extra cooling should
be present in the NSs because of the soft photons emitted by the  
surface. This seems indeed true for those systems in which a high energy
cutoff has been observed, although some LMXBs do not show any cutoff in 
their spectra up to $\sim 100$ keV (Barret et al. 1991; Harmon et al. 1996; 
Piraino et al. 1999).

This similarity with the spectra of the BHs suggests that the same 
emission mechanism and geometry operate in NSs.
In this scenario the soft excess comes from the accretion disk.
The comptonized component originates in a hot and optically
thin cloud (corona), probably placed between the neutron star and the 
accretion disk. 
A fraction of this component is reprocessed by the surface of the 
accretion disk arising
the reflection bump and the iron K$_\alpha$ line. The weakness of
these components suggests that the reflector subtends a small solid angle
($\Omega/2\pi < 0.3$) as seen from the corona. This is possible if the 
disk is truncated and its inner part is absent, or if the inner accretion 
disk is highly ionized (but see Done \& $\dot{\rm Z}$ycki 1999).  
In this geometry, a decrease of the inner radius of the disk causes
an increase of the solid angle subtended by the reflector and a steepness
of the power law.  This could be responsible of the observed correlation 
between the photon index of the power law component and the reflection 
amplitude.

An Advection Dominated Accretion Flow (ADAF, see Narayan \& Yi 1995 for a 
description) has been proposed as the origin of the scattering corona in 
NSs as in BHs (see Barret et al. 1999 for a discussion on this argument). 
However there is an important difference between NSs and BHs, 
because the neutron stars have a solid surface and the advected energy
should be released at the surface or in a boundary layer.  This extra soft 
emission would enhance the Compton cooling and result in softer spectra. 
Barret et al. (1999) suggest that the boundary layer is optically thin and 
its emission is hard, even if a reprocessed component from the neutron star 
surface could also be present.

In this paper we concentrate on the broad band (0.1--100 keV) spectral
analysis performed on a BeppoSAX observation of MXB 1728--34 (4U 1728--34).
MXB 1728--34 is a Low Mass X-ray Binary (LMXB), belonging to the class of
atoll sources (Hasinger and van der Klis 1989).
The optical counterpart has not been 
identified yet, due to the high optical extinction in the galactic center
direction. The distance to this source is poorly known. It is probably
between 4 and 14 kpc (Grindlay \& Hertz 1981).

Observations with the Rossi X-Ray Timing Explorer (RXTE) have shown 
kilohertz quasi-periodic oscillations (kHz QPO) in the persistent emission 
of MXB 1728--34, with frequencies ranging from a few hundred Hz up to 
$\sim 1200$ Hz (Strohmayer et al. 1996).  Usually two kHz peaks 
are simultaneously observed, with a nearly constant difference between 
their frequencies.  If the frequency upper kHz QPO is interpreted as the
Keplerian frequency at the inner rim of the accretion disk ({\it e.g.} Miller,
Lamb, \& Psaltis 1998, Stella \& Vietri 1999), this implies that the disk can
arrive very close to the neutron star in these systems.
MXB 1728--34 shows frequent type-I X-ray bursts (Basinska et al. 1984). 
Recently bursts from this source were observed by RXTE 
(Strohmayer et al. 1997), and nearly coherent oscillations were discovered
during the rising phase of the bursts, 
at a frequency of 363 Hz. The oscillation amplitude 
decreased as the X-ray burst flux increased. These oscillations likely 
result from spin modulation of a not uniform X-ray flux produced by the 
thermonuclear burning that causes the burst. 
The frequency separation
between the two simultaneous kHz QPOs is similar to the frequency of the
burst oscillations, suggesting a beat frequency mechanism as the origin
of the kHz QPOs (Strohmayer et al.  1996, Miller et al. 1998).  
However precise measures of the peak separation
between the kHz QPOs in this source have shown that it is always smaller 
than 363 Hz, and decreases significantly at higher inferred accretion 
rates (M\'endez \& van der Klis 1999).

The spectrum during the bursts is usually fitted by a blackbody. 
In the rising phase the blackbody temperature increases from $\sim 1$ 
to $\sim 2.5$ keV (Foster et al. 1986; Day \& Tawara 1990), and, during the 
burst decay, decreases  down to $\sim 1$ keV (Foster et al. 1986).
There is some evidence of the presence of a hard tail during 
bursts (Foster et al. 1986). From the burst spectrum the radius of
the emitting region can be derived and compared with the radius of the
neutron star. Usually values well below 10 km are found (van Paradijs 1978,
see also Basinska et al. 1984 and references therein), {\it i.e.} below the
lowest radius allowed for a 1.4 $M_\odot$ neutron star. 
Foster et al. (1986) showed that the measured color temperature of the
blackbody can be higher than the effective temperature. Using appropriate
correcting factors they derive a radius of the emitting region which is
consistent with current equations of state, and a distance to the source
of 4.3 kpc.

The persistent spectrum of MXB 1728--34 in the 1--20 keV energy range, 
observed by {\it Einstein} MPC, was fitted by a thermal bremsstrahlung 
with a temperature of $\sim 18$ keV (Grindlay and Hertz 1981).
The spectrum in the same energy range was also obtained from SAS 3 data
(Basinska et al. 1984).
During the SAS 3 long observation ($\sim 48$ days) the source intensity 
(in the 1--20 keV energy range) was found to vary by a factor $\sim 2$. 
Accordingly the temperature associated to the spectrum was found
to vary between 4 and 11 keV, positively correlated with the X-ray flux.
The hard energy spectrum (30--200 keV), observed by SIGMA telescope
(Claret et al. 1994), was also fitted by a thermal bremsstrahlung but
with a higher temperature ($\sim 38$ keV). The 30--200 keV flux 
was $\rm \sim 7 \times 10^{-10}\;erg\;cm^{-2}\;s^{-1}$.

\section{Observations}

The Narrow Field Instruments (NFI) on board BeppoSAX satellite
(Boella et al. 1997) observed MXB 1728--34 on 1998 August 
23 and 24, for a total exposure time of 24 ks.  
The NFIs are four co-aligned instruments which cover more
than three decades of energy, from 0.1 keV up to 200 keV, with good
spectral resolution in the whole range. LECS (operating in the range 0.1--10
keV) and MECS (1--11 keV) have
imaging capabilities with a Field of View (FOV) of $20'$ and $30'$ radius 
respectively. We selected the data for scientific analysis 
in circular regions of $8'$ and $4'$ radii for the LECS and MECS 
respectively centered on the source. The background subtraction
was obtained using blank sky observations, in which we extracted the 
background spectra in regions of the FOV similar to those used for the
source. MXB 1728--34 is near the Galactic plane and in this case simultaneous
backgrounds should be used (Parmar et al. 1999). However MXB 1728--34 is an 
intense source and the background contributes to less than 1\% and 0.3\% of
the total count rate in the LECS and MECS respectively. Therefore we preferred
to use the standard method for background subtraction in this case. 
HPGSPC (7--60 keV) and PDS (13--200 keV) have no imaging 
capabilities, because the FOVs, of $\sim 1^\circ$ FWHM, are delimited 
by collimators. The background subtraction for these instruments was 
obtained using the off-source data accumulated during the rocking of the 
collimators.

During the persistent emission the MECS light curve
shows little variability of the intensity, the count rate ranging between 25 
and 30 counts/s. 
We divided the MECS data in two energy bands, soft (1.8--4 keV) and hard 
(4--10 keV), and calculated the corresponding hardness ratio.
This hardness ratio is constant implying no spectral variations during
the BeppoSAX observation.
The average observed flux of the source (in the 0.1--100 keV energy range) 
during the persistent emission is
$3.6 \times 10^{-9}$ ergs cm$^{-2}$ s$^{-1}$, which corresponds to an 
unabsorbed luminosity of $1.2 \times 10^{37}\ D_4^2$ ergs/s, where $D_4$
is the distance to the source in units of 4 kpc.

During the BeppoSAX observation of MXB 1728--34, three type-I bursts
are present. The duration of these bursts is between 10 and 15 s, 
considering the start and end 
times of each burst when the flux is 10\% of the peak above the persistent 
emission level. The peak count rate (measured by the MECS) is around 
500 counts/s for the first and the second burst, and $\sim 300$ counts/s 
for the third one. 
Figure 1 shows the MECS light curve (upper panel), in the energy band 1.8--10 
keV and with the persistent level subtracted, of the first burst and the 
corresponding hardness ratio (4--10 keV/1.8--4 keV, lower panel). 
The hardness ratio seems to decrease from $\sim 2.5$ to $\sim 1$ during the 
burst decay. This means that the spectrum softens, as expected by the decay 
of the blackbody temperature during the burst.
We obtain similar results for the other two bursts.

\section{Spectral Analysis}
The energy ranges used in the spectral analysis for each NFI are: 
0.12--4 keV for the LECS, 1.8--10 keV
for the MECS, 8--30 keV for the HPGSPC and 15--100 keV for the PDS.
Different normalizations of the four NFIs are considered by including in 
the model normalizing factors, fixed to 1 for the MECS and kept free for 
the other instruments. 

\subsection{Spectral Analysis of the Burst Emission}

We used the MECS and PDS data to produce the spectra during the bursts,
considering the start and end times of each burst when the rate in the MECS 
is 10\% of the peak above the persistent emission level.
LECS and HPGSPC data were not used in this case because of the low statistics 
due to the short duration.  With a preliminary fit
we verified that the spectra of the three bursts presented the same  
shape. Therefore, to increase the statistics, we summed the three spectra 
obtained from the MECS. In the PDS, only the second and third bursts were 
observed and summed together. During the first burst, which falls at the
beginning of a BeppoSAX orbit, the PDS was switched off.
From the obtained burst spectrum we subtracted the spectrum of the persistent 
emission, and the result is shown in Figure 2 (upper panel).
We fitted this spectrum with a 
blackbody modified by interstellar absorption from cold matter.
Because of the lack of data at energies below 1.8 keV, we fixed the 
value of the equivalent hydrogen column N$_{\rm H}$ at $2.5 \times 10^{22}$ 
cm$^{-2}$, in agreement with the value obtained from the fit of the BeppoSAX 
spectrum during the persistent emission (Table 2, model 4) and with 
previous results (Hoffman et al. 1979, Grindlay \& Hertz 1981, 
Foster et al. 1986).
In Table 1 we report the best fit values of the parameters and in Figure 2 
(lower panel) the residuals (in units of $\sigma$) with respect to the 
best fit model.  A hard excess between 20 and 30 keV is visible in these 
residuals with low statistical significance. This hard excess is still present
when we restrict the MECS data to only the second and third bursts, that
are simultaneously observed by MECS and PDS. Indeed, fitting only the PDS 
data to a blackbody we obtain a higher temperature, between 2.3 and 2.9 keV.

\subsection{Spectral Analysis of the Persistent Emission}

The broad band (0.1--100 keV) spectrum during the persistent
emission was obtained excluding intervals of $\sim 100$ s around each
burst, and is shown in Figure 3 (upper panel).
A simple model (consisting of a single component, like a thermal 
bremsstrahlung with photoelectric absorption by cold matter)
was not sufficient to fit the spectrum in the whole energy band. 

We tried several models and we obtained a good representation of the spectrum 
using a soft emission that could be equivalently modeled by a multi-color 
disk blackbody ({\it diskbb} in XSPEC, Mitsuda et al. 1984) or a blackbody, 
a comptonized component fitted with the {\it comptt} model (described in 
Titarchuk, 1994), and two gaussian emission lines. 
The results of this fits are shown in Table 2 (model 1 and 2 respectively).
The comptonized spectrum seems to be better fitted by the {\it thComp} model 
(see the appendix of Zdziarski et al. 1996 for a description of this model). 
The results of this
fit are shown in Table 2, model 3 with the soft excess described by a 
multi-color disk blackbody, and model 4 with the soft excess described by
a blackbody.  The unfolded spectrum, together with the spectral components 
used to fit the data, are shown in Figure 4.

The soft emission could be described by a blackbody with a temperature of
$\sim 0.5$ keV or a multi-color disk blackbody with an inner temperature 
$\sim 0.8$ keV.
As it is evident from Table 2, the values of the parameters characterizing
the other components do not vary significantly adopting
a multi-color disk blackbody or a blackbody to fit the soft excess.
For the comptonized spectrum, we obtain an electron temperature of 
$\sim 7-11$ keV and an optical depth of $\tau \sim 5$ for a spherical 
scattering cloud, with a temperature of the seed photons $\sim 1.5$ keV. 
We used a broad gaussian to fit an iron K$_\alpha$ line at $\sim 6.7$ keV,
with equivalent width of 50--70 eV.
We also used a narrow gaussian line to fit an excess in the residuals around 
1.7 keV. The addition of a gaussian emission line at this energy is 
statistically significant at more than 99.99\% confidence level. 
The residuals with respect to model 4 are shown in Figure 3 (lower panel).
We also tried the presence of an iron edge at 7--8 keV, but the addition of
this component to the model does not improve the fit.

The addition of a reflection component ({\it pexriv} in XSPEC, Magdziarz
\& Zdziarski 1995) does not improve significantly the fit. 
The upper limit on the reflection amplitude, corresponding to the solid 
angle (in units of $2 \pi$) that the reflector subtends as viewed from the 
corona, is $\sim 0.7$, using an inclination angle 
$\cos i = 0.5$. However the {\it pexriv} model does not contain the iron 
emission line that should be present in the reflection component.
Therefore we tried a reflection model (that we shall call {\it fe-refl})
in which the iron line is self-consistently calculated for the given 
ionization state, temperature, spectral shape and abundances
(this model is described in detail in $\dot{\rm Z}$ycki et al. 1998). 
Excluding the gaussian line at $\sim 6.7$ keV and adding this self-consistent 
reflection model (with the inclination angle $\cos i$ fixed to 0.5), 
we obtain a fit equivalent to that shown in Table 2 (model 4), with
$\chi^2 = 663$ for 614 degrees of freedom.
The observed iron line requires, to be fitted, reflection from a ionized
disk (with $\xi = L_{\rm X}/n_{\rm e} r^2 \sim 280$) and a reflection 
amplitude $\Omega/2\pi = 0.15 \pm 0.08$.

\section{Discussion}

We performed a spectral analysis of the bursts observed by BeppoSAX 
in the 1.8--70 keV energy range.  We fitted the averaged spectrum during the
bursts to a blackbody
with a temperature $\sim 2$ keV.  The total luminosity of this blackbody 
is $3.6 \times 10^{37}\ D_4^2$ ergs/s, where $D_4$ is the distance to the 
source in units of 4 kpc.
This corresponds to a radius of the emitting region of $\sim 4.8\ D_4$ km. 
Because the most probable distance to the source is not much different
from 4 kpc (see Foster et al. 1986, and below), this radius is rather 
low to be identified with
the neutron star radius. However the color temperature of the spectrum
can be higher than the effective temperature by a factor 1.4 (Foster et al. 
1986). Considering this correcting factor we find an effective radius
of the emitting region of $R_{\rm eff} \simeq
9.4\ D_4$ km, that is comparable with the radius of a neutron star.

The distance to the source is not well 
known and is estimated between 4 and 14 kpc (Grindlay \& Hertz 1981). 
However the bursts of this source can be used to infer the distance,
as shown by Basinska et al. (1984). They found that the maximum
burst flux ($F_{\rm max}$) and the integrated burst flux ($E_{\rm b}$)
are strongly correlated at low flux levels. For high flux levels $F_{\rm max}$
saturates at the value of $7 \times 10^{-8}$ ergs cm$^{-2}$ s$^{-1}$ 
(calculated in the 1--20 keV energy range). This is therefore a critical 
luminosity and might be interpreted as the Eddington limit luminosity.
The bursts we see in MXB 1728--34 during the BeppoSAX observation are far 
from this saturation level. The nearest to saturation is the first burst,
for which $E_{\rm b} = 3.3 \times 10^{-7}$ ergs cm$^{-2}$ and
$F_{\rm max} = 4.5 \times 10^{-8}$ ergs cm$^{-2}$ s$^{-1}$ (corresponding to 
$\sim 64\%$ of the critical luminosity level), both calculated in the 1--20
keV energy range and with the persistent level subtracted as in Basinska et 
al. (1984). Using the fit parameters of Table 1, we can correct the 
maximum burst flux for absorption and extrapolate it in the 0.1--100 keV energy
range. This corresponds to a total maximum luminosity of 
$6.2 \times 10^{36}\ D^2$ ergs/s, where D is the distance in kpc.
In the interpretation above, this should be equal to 64\% of the Eddington
luminosity, that is $L_{\rm E} = 2.5 \times 10^{38}$ ergs/s for a 1.4 
$M_{\odot}$ (van Paradijs \& McClintock 1994), giving a distance to the source
of $\sim 5.1$ kpc. This is similar to previous estimations based on the bursts
({\it e.g.} 4.3 kpc, Foster et al. 1986, and 4.2 kpc, van Paradijs 1978).

A rather complex model is needed to fit the persistent spectrum. The 
continuum consists of a soft emission that could be described by a
blackbody or a multi-color disk blackbody plus a comptonized spectrum.
This model is similar to that used for other type-I X-ray bursters,
such as 1E 1724--3045 (Guainazzi et al. 1998, Barret et al. 1999) and 
KS 1731--260 (Barret et al. 1999). For these sources and MXB 1728--34, the 
electron temperature of the scattering region is between 3 and 28 keV and the
optical depth between 3 and 10, with higher temperatures corresponding
to lower optical depths, as expected.  In the following we will describe 
the results we obtained for MXB 1728--34 in more detail. 

The value of the equivalent absorption column $N_{\rm H}$ is 
2.5--3 $\times 10^{22}$ cm$^{-2}$, depending on the model used to
fit the soft emission (blackbody or disk multi-temperature blackbody).
If the distance of 5.1 kpc inferred from the bursts is correct, this 
implies a visual extinction of $A_{\rm v} = 7.8 \pm 1.7$ for MXB 1728--34
(calculated from Hakkila et al. 1997). Using the observed correlation
between visual extinction and absorption column (Predehl \& Schmitt 1995)
we find $N_{\rm H} = (1.4 \pm 0.3) \times 10^{22}$ cm$^{-2}$, which is
around half the value we find from the spectral fit. Large values of the
absorption column, between 1.5 and 3.5 $\times 10^{22}$ cm$^{-2}$ and even
larger, were also found in previous observations of MXB 1728--34 (see 
Basinska et al. 1994). Therefore there is an excess of attenuation of the 
soft emission that has to be caused by matter close to the X-ray source, 
{\it e.g.} by the outer cold regions of an accretion disk or corona.

Using a blackbody to describe the soft emission, we get a temperature of
$\sim 0.5$ keV. The luminosity of this component is $\sim 3 \times 10^{36}\
D_4^2$ ergs/s. This implies a radius of the spherical emitting region of 
$\sim 15\ D_4$ km. Indeed the distance to the source could be
higher than the 4 kpc (Grindlay \& Hertz 1981) and therefore the luminosity
and the radius of the blackbody emission region could be higher. 
For instance, adopting a distance of 5.1 kpc, the calculated radius is 
$\sim 20$ km.  Therefore the radius of the blackbody emitting region could be
rather large to be identified with the neutron star surface.
Using a multi-color disk blackbody to model the soft excess we obtain an 
inner disk temperature of $\sim 0.8$ keV, and a lower limit for the inner 
radius of the disk of $R_{\rm in} \sqrt{\cos i} \simeq 8\ D_4$ km.
This source does not show dips in the light curve and therefore the  
inclination of the normal to the plane of the disk with respect to the 
line of sight should be less than $60^\circ$. Assuming an inclination angle
$i$ of $60^\circ$ we obtain $11\ D_4$ km for the inner radius of 
the accretion disk. 
Moreover the simple multi-color disk blackbody could be not appropriate 
to describe the emission of accretion disks in X-ray binaries, because the
electron scattering will modify the spectrum (Shakura \& Sunyaev 1973; 
White, Stella \& Parmar 1988). In this case, the measured color temperature
is related to the effective temperature of the inner disk 
$T_{\rm col} = f T_{\rm eff}$, where $f$ is the spectral hardening factor.
The factor $f$ has been estimated by Shimura \& Takahara (1995) to be
$\sim 1.7$ for a luminosity $\sim 10\%$ of the Eddington limit, with a
little dependence on the mass of the compact object and the radial position.
Applying this correction to the values of $T_{\rm in}$ and 
$R_{\rm in} \sqrt{\cos i}$ reported in Table 2, we obtain an effective 
temperature of $\sim 0.5$ keV, which corresponds to an inner radius 
$R_{\rm eff} \sqrt{\cos i} = f^2 R_{\rm in} \sqrt{\cos i} \simeq 20\ D_4$ km.

In conclusion, we identify the soft component with the emission 
from the accretion disk, although we obtain a slightly better fit 
using the blackbody instead of the multi-color disk blackbody, 
especially when we describe the Comptonized spectrum with {\it thComp}
(compare model 3 and model 4 in Table 2). In the hypothesis that the
blackbody represents the emission from the inner accretion disk, we can
calculate the inner radius of the disk from this component, in order to
compare the result with the estimate obtained using the multi-color disk 
blackbody for the soft excess. The blackbody luminosity is then the total
potential energy that has been released at $R_{\rm in}$:
\begin{equation}
L_{\rm BB} = \frac{G M \dot{M}}{2 R_{\rm in}}
\end{equation}
where we assumed a Keplerian, geometrically thin, optically
thick accretion disk.
We attribute the measured temperature of the blackbody to the maximum
temperature in the disk, that is reached close to the inner radius at
$R = 49/36$ $R_{\rm in}$ (see {\it e.g.} Frank, King, \& Raine 1985).
\begin{equation}
T_{\rm max} = 0.488 \left(\frac{3 G M \dot{M}}{8 \pi \sigma R_{\rm in}^3} 
\right)^{1/4}
\end{equation}
where we assumed that the disk is truncated at $R_{\rm in}$ and that the
zero-torque condition can be applied at $R_{\rm in}$. Using these two
equations we can find the inner radius of the disk as a function of the 
measured temperature and luminosity of the blackbody:
\begin{equation}
R_{\rm in} = 3.69 T_{\rm keV}^{-2} L_{37}^{1/2} \; {\rm km}
\end{equation}
Considering the values reported in Table 2 (model 4), we obtain 
$R_{\rm in} = 6.3\ D_4$ km for MXB 1728--34, that becomes $18\ D_4$ km 
considering a spectral hardening factor of $\sim 1.7$. This value is 
similar to the result obtained with the multi-color disk blackbody.

The second component of our model is a comptonized spectrum, produced in
a hot ($k T_{\rm e} \sim 7-11$ keV) region of spherical shape and
moderate optical depth ($\tau \sim 5$) surrounding the neutron star. 
This comptonized spectrum requires
soft seed photons at a temperature of $\sim 1.4-1.5$ keV, that is 
significantly higher than the temperature of the disk.
The emission from the neutron star surface and/or
from an optically thick or thin boundary layer (see the discussion in Barret 
et al. 1999 on the possible existence of an optically thin boundary layer) 
could then account for the seed photons subsequently comptonized in the
surrounding corona. This could explain why we do not observe directly the
emission from the neutron star and is in agreement with our identification
of the blackbody as the emission from the accretion disk.

Adopting the geometry described above one would expect a Compton reflection
by the accretion disk of the hard spectrum 
emitted by the corona. This would produce a bump in the spectrum, 
that can be described by the {\it pexriv} model. This reflection is not 
significantly detected in MXB 1728--34 (nor in the similar sources 
1E 1724--3045, Guainazzi et al. 1998, Barret et al. 1999, and KS 1731--260, 
Barret et al. 1999). A possible reason could be the low temperature
of the spectrum emitted by the corona. Because most of the hard photons have 
energies below 10 keV, they will be photoelectric absorbed by the matter 
in the disk, and re-emitted at the temperature
of the disk, rather than Compton scattered. This explains why we 
obtained a rather high upper limit in the reflection amplitude ($\sim 0.7$)
using {\it pexriv}. It is probable that the reflection component can be 
observed in a harder state of this source. 

This argument could also be applied to KS 1731--260, whose spectrum is 
indeed soft, with an electron temperature of $\sim 2.8$ keV 
(Barret et al. 1999), but shows some difficulties when applyed to 
1E 1724--3045, whose spectrum is harder with an electron temperature 
$\sim 28$ keV (Guainazzi et al. 1998, Barret et al. 1999). In this case
Barret et al. (1999) suggest that the inclination of the source could be high
and this could reduce the reflection component. We note that 
up to date the reflection component has been detected only in sources with
very hard spectra, such as GS 1826--238 (with a cutoff energy $\sim 90$ keV,
Barret et al. 1999), SLX 1735--269 (with a cutoff energy $\sim 200$ keV,
Barret et al. 1999), 4U 0614+09 (with a cutoff energy $> 200$ keV,
Piraino et al. 1999) and 4U 1608--52 (with a cutoff energy at 300 keV,
Yoshida et al. 1993).

The irradiation of the accretion disk by the X-ray primary spectrum
should also produce an iron emission line at $\sim 6.4$ keV.  
Therefore we can better constrain the parameters of the reflection component
with the {\it fe-refl} model, which includes a self-consistent 
calculation of this line. Using such a model, instead of the gaussian, 
to fit the iron line, we find that the reflector has to be ionized 
($\xi \sim 280$) and we derive an upper limit on the reflection 
amplitude of $\sim 0.2$, much stringent than the value found using 
{\it pexriv}.

Two emission lines, at 6.7 keV and 1.66 keV, are needed to fit 
the spectrum of MXB 1728--34. The line at $\sim 6.7$ keV is interpreted as 
emission from highly ionized iron.  A broad emission line at 6.7 keV is 
compatible with being emitted in a strongly ionized region such as the 
corona or the inner part of the disk. The second 
emission line, at $\sim 1.66$ keV, is compatible with the radiative 
recombination emission from Mg XI. These X-ray emission lines at low 
energies likely arise in a photoionized corona (Liedahl et al. 1992, see
White, Nagase \& Parmar 1995 as a review).

\section{Conclusions}

We analysed data from a BeppoSAX observation of MXB 1728--34 performed in
1998 August between 23 and 24. Three type-I X-ray bursts were present 
during this observation. The spectrum during the bursts is fitted by a
blackbody with a temperature $\sim 2$ keV. We calculated the corresponding
effective radius of the blackbody emitting region, that is $R_{\rm eff}
\simeq 9.4\ D_4$ km, comparable with the neutron star radius.
Assuming that the peak flux {\it vs.} fluence relationship of the bursts 
saturates at the Eddington limit luminosity, we find a distance
to the source of 5.1 kpc. 

The spectrum during the persistent emission is described
by a blackbody with a temperature of $\sim 0.5$ keV, a comptonized
spectrum, and two gaussian emission lines at 1.66 keV and 6.7 keV 
respectively. In the hypothesis that the blackbody is emitted by the 
accretion disk we estimate an inner disk radius of $\sim 20$ km.
The comptonized spectrum is probably produced in a spherical, hot 
($\sim 10$ keV) region of moderate optical depth ($\tau \sim 5$) 
surrounding the neutron star. The soft seed photons for the Comptonization,
with a temperature $\sim 1.5$ keV, probably come from the neutron star
surface and/or boundary layer. The presence of a reflection component
is not detected with high statistical significance. To constrain the
reflection parameters we used a self-consistent model that also calculates 
the iron line produced in the reflection. This is important because
MXB 1728--34 was in a soft state during the BeppoSAX observation and
the low temperature of the primary emission significantly reduce the reflection
bump between 20 and 40 keV. Using this self-consistent model to fit the iron 
line, we find that the reflector is ionized and the reflection amplitude 
is $\sim 0.2$. Alternatively the iron line could be emitted in a 
strongly ionized corona. Further observations are needed to 
clarify the origin of the iron line.

\acknowledgments
The authors want to thank P.T. $\dot{\rm Z}$ycki and C. Done for kindly
supplying their reflection model {\it fe-refl} and the Comptonization model
{\it thComp}, and for useful discussions.
This work was supported by the Italian Space Agency (ASI), by the Ministero
della Ricerca Scientifica e Tecnologica (MURST).



\section*{TABLES}

\begin{table}[th]
\small
\begin{center}
\begin{tabular}{l|c} 
\tableline \tableline
    Parameter                     & Value \\ 
\tableline
$N_{\rm H}$ $\rm (\times 10^{22}\;cm^{-2})$ & 2.5 (frozen) \\
$k T_{\rm BB}$ (keV)               & $1.90 \pm 0.06$ \\
N$_{\rm BB}$                   & $0.223 \pm 0.009$ \\
$\chi^2$/d.o.f.                   & 233/184  \\
\tableline
\end{tabular}
\caption{Results of the fit of the averaged burst spectrum (see text) with 
a blackbody in the energy band 1.8--70 keV.
Uncertainties are at the 90\% confidence level for a single parameter.
Blackbody normalization (N$_{\rm BB}$) is in units of $L_{39}/D_{10}^2$, 
where $L_{39}$ is the luminosity in units of $10^{39}$ ergs/s and $D_{10}$ 
is the distance in units of 10 kpc.} 
\label{tab1}
\end{center}
\end{table} 

\clearpage
\begin{table}[th]
\small
\begin{center}
\begin{tabular}{l|c|c|c|c} 
\tableline \tableline
 Parameter        & Model 1 & Model 2 & Model 3 & Model 4 \\
               & Comptt + Diskbb & Comptt + BB & thComp + Diskbb  
                                                 & thComp + BB   \\ 
\tableline
$N_{\rm H}$ $\rm (\times 10^{22}\;cm^{-2})$ & $3.1 \pm 0.1$ & $2.59 \pm 0.09$ &
$3.14 \pm 0.10$ & $2.674 \pm 0.098$  \\
$k T_{\rm in}$ (keV)           & $0.82 \pm 0.04$ & -- & $0.753 \pm 0.031$ 
& --   \\
$R_{\rm in}$ $\sqrt{\cos i}$ (km) & $6.96 \pm 0.72$ & -- & $8.07 \pm 
0.89$ & --   \\
$k T_{\rm BB}$ (keV)    		  & -- & $0.583 \pm 0.014$ & -- &
$0.551 \pm 0.014$  \\
N$_{\rm BB}$          & -- & $1.872 \pm 0.062$ & -- &
$1.670 \pm 0.073$   \\
$k T_0$ (keV)			  & $1.531 \pm 0.048$ & $1.419 \pm 0.033$ & 
$1.473 \pm 0.050$ & $1.388 \pm 0.037$ \\
$k T_{\rm e}$ (keV)                & $7.07^{+1.2}_{-0.79}$ & $6.31^{+0.74}_{-0.55}$ &
$11.3^{+5.1}_{-2.4}$ & $9.4^{+2.5}_{-1.5}$  \\
$\tau$				  & $4.56 \pm 0.70$ & $5.26 \pm 0.60$ & -- & 
--  \\
Photon Index                      & -- & -- & $3.13 \pm 0.17$ &
$2.98 \pm 0.13$  \\
N$_{\rm comp}$ $(\times 10^{-2})$  & $5.58 \pm 0.97$ & $6.91 \pm 0.89$ &
$4.41 \pm 0.33$ & $5.10 \pm 0.28$   \\
$E_{\rm Fe}$ (keV)                    & $6.68 \pm 0.13$ & $6.66 \pm 0.12$ & 
$6.72 \pm 0.13$ & $6.74 \pm 0.13$   \\
$\sigma_{\rm Fe}$ (keV)               & 0.5 (frozen) &  0.5 (frozen) &
0.5 (frozen) & $0.34^{+0.20}_{-0.13}$   \\
$I_{\rm Fe}$ $(\times 10^{-3})$       & $1.94 \pm 0.47$ & $1.46 \pm 0.46$ & 
$2.01 \pm 0.45$ & $1.34^{+0.57}_{-0.41}$   \\
EW$_{\rm Fe}$ (eV)		      & $69 \pm 17$ & $51 \pm 16$ &
$72 \pm 16$ & $48 \pm 18$  \\
$E_{\rm LE}$ (keV)                    & $1.659 \pm 0.031$ & $1.670 \pm 0.021$ & $1.656^{+0.030}_{-0.016}$ & $1.664 \pm 0.027$   \\
$\sigma_{\rm LE}$ (keV)               & $< 0.05$ & $< 0.05$ & $< 0.06$ &
$< 0.05$   \\
$I_{\rm LE}$ $(\times 10^{-2})$       & $1.49 \pm 0.53$ & $1.18 \pm 0.37$ & 
$1.56 \pm 0.60$ & $1.24 \pm 0.40$   \\
EW$_{\rm LE}$ (eV)			  & $41 \pm 14$ & $46 \pm 14$ &
$42 \pm 16$ & $47 \pm 15$  \\
$\chi^2$/d.o.f.                   & 687/614 & 681/614 & 676/614 & 658/613 \\
\tableline
\end{tabular}
\caption{Results of the fit of the averaged spectrum, in the energy band 
0.12--100 keV, during the persistent 
emission with a multi-color disk blackbody (Diskbb) or a blackbody (BB), a 
comptonized spectrum modeled by Comptt or thComp (see text), and two
gaussian emission lines.
Uncertainties are at the 90\% confidence level for a single parameter.
In the Diskbb model the 
inner radius is calculated for a distance to the source of 4 kpc.  
The blackbody normalization (N$_{\rm BB}$) is
in units of $L_{37}/D_{10}^2$, where $L_{37}$ is the luminosity in 
units of $10^{37}$ ergs/s and $D_{10}$ is the distance in units of 
10 kpc.  $k T_0$ is the temperature 
of the soft seed photons for the Comptonization, $k T_{\rm e}$ is the 
electron temperature, and $\tau$ is the optical depth of the spherical 
scattering cloud. Comptt normalization is in XSPEC units and thComp 
normalization is in photon cm$^{-2}$ s$^{-1}$ at 1 keV.
$\sigma_{\rm Fe}$ was fixed to 0.5 keV when its value could not be 
derived from the fit.
$E_{\rm LE}$ is the centroid energy of the low energy emission
line.} 
\label{tab2a}
\end{center}
\end{table}

\clearpage
 
\section*{FIGURES}
\vskip 3.7cm

\begin{figure}[h]
\centerline
{\psfig
{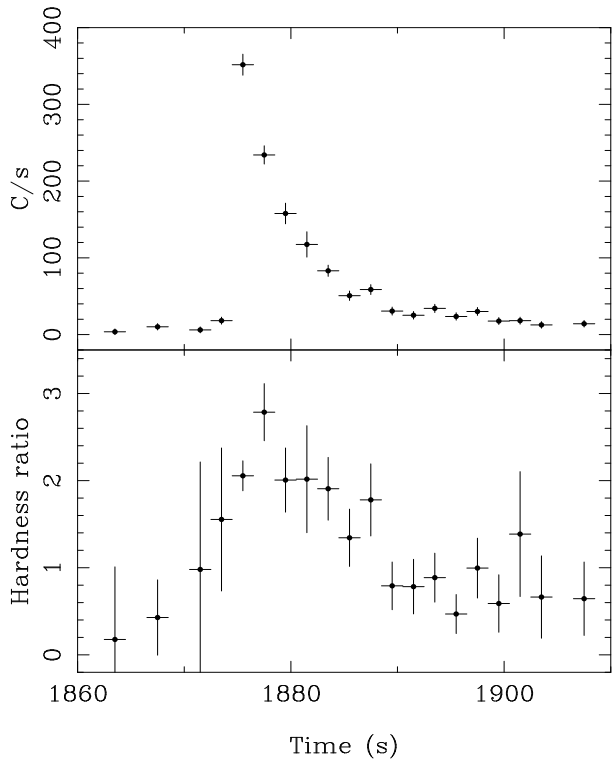}}
\caption{Upper panel: Light curve in the MECS energy range 
(1.8--10 keV) of the first burst observed by BeppoSAX. The persistent 
emission level was subtracted. The bin time is 2 s.
Lower panel: The corresponding hardness ratio, defined as the 
ratio of the counts in the energy range 4--10 keV to the counts in the
energy range 1.8--4 keV. In both the energy bands the corresponding 
persistent emission level was subtracted.}
\label{fig1}
\end{figure}

\begin{figure}[t!]
\centerline
{\psfig
{figure=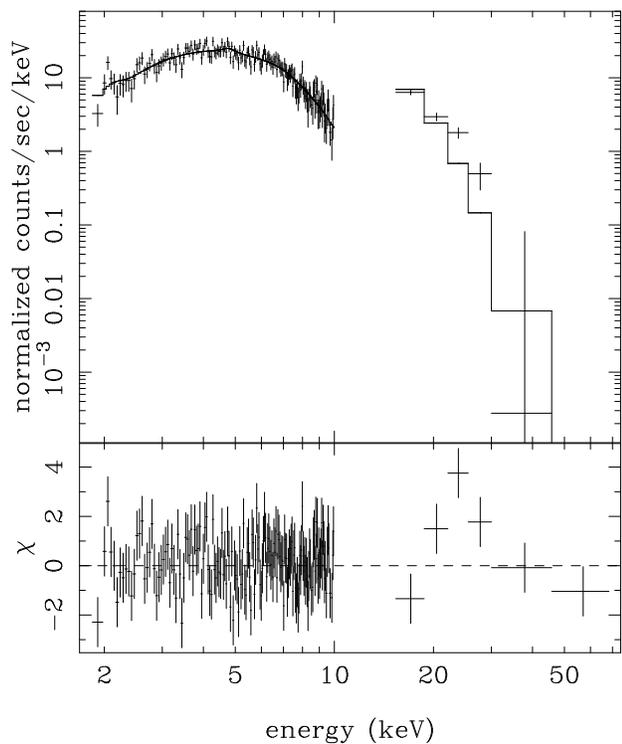,height=12.0cm,width=9.0cm}}
\caption{Upper panel: Averaged count spectrum during the bursts 
(see text) in the MECS
(1.8--10 keV) and PDS (15--70 keV) ranges. Lower panel: Residuals in units of
$\sigma$ with respect to a blackbody model. A hard tail is marginally
evident between 20 and 30 keV.}
\label{fig2}
\end{figure}

\begin{figure}[t!]
\centerline
{\psfig
{figure=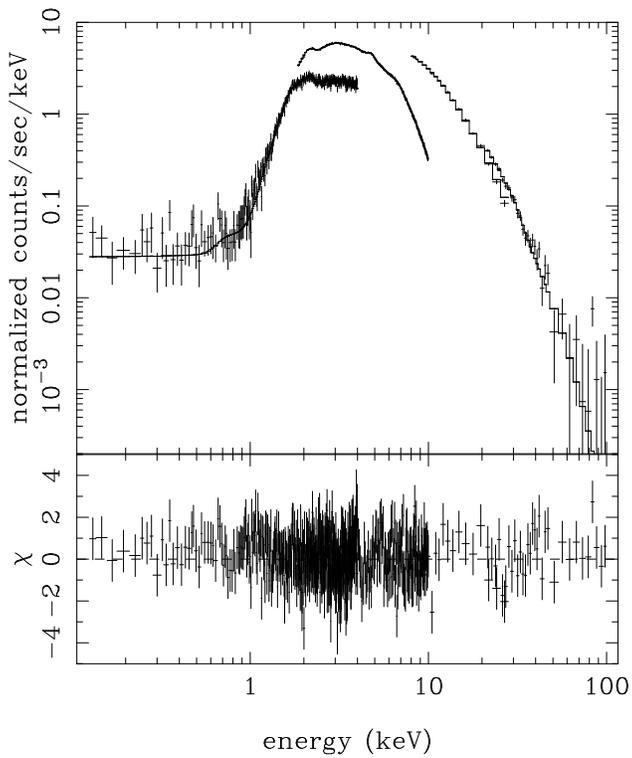,height=12.0cm,width=9.0cm}}
\caption{Upper panel: Broad band 
spectrum of MXB 1728--34 (0.1--100 keV) during the persistent emission,
and the best fit model reported in Table 2 (model 4). 
Lower panel: Residuals in units of $\sigma$ with respect this model.}
\label{fig3}
\end{figure}

\begin{figure}[t!]
\centerline
{\psfig
{figure=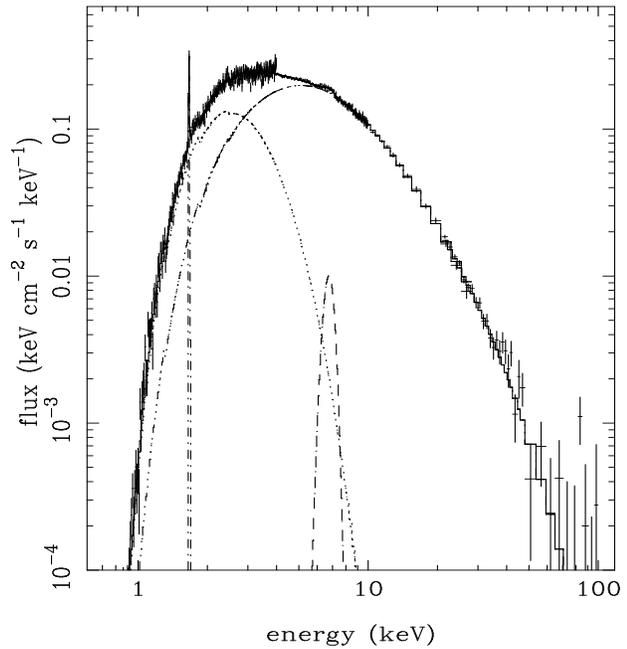,height=10.0cm,width=9.0cm}}
\caption{Unfolded spectrum of the persistent emission 
and the best fit model (model 4 of Table 2), shown in this figure as a
solid line. The single components of this model are also shown, namely
the blackbody (dotted line), the comptonized spectrum ({\it thComp} model,
dot-dot-dot-dashed line), and two gaussian emission lines at $\sim 1.7$ keV 
(dot-dashed line) and $\sim 6.7$ keV (dashed line) respectively.}
\label{fig4}
\end{figure}

\end{document}